\newcommand{\age}{\ \raisebox{-.3ex}{$\stackrel{>}{\scriptstyle \sim}$}\ }

\documentclass{mn2e}

\input psfig

\title[T Tauri disc evolution]
	{Dispersion in the lifetime and accretion rate \\ of T Tauri discs}
	
\author[P.J. Armitage, C.J. Clarke \& F. Palla]
       {Philip J. Armitage$^{1,2}$\thanks{email: {\tt pja@jilau1.colorado.edu}}, 
       Cathie J. Clarke$^3$ and Francesco Palla$^4$ \\
	$^1$JILA, University of Colorado, 440 UCB, Boulder CO 80309-0440, USA \\
	$^2$Department of Astrophysical and Planetary Sciences, University of Colorado, 
	    Boulder CO 80309-0391, USA \\ 
	$^3$Institute of Astronomy, Madingley Road, Cambridge CB3 0HA \\
	$^4$Osservatorio Astrofisico di Arcetri, Largo E. Fermi 5, 50125 Firenze, Italy}	

\begin{document}

\maketitle

\begin{abstract}
We compare evolutionary models for protoplanetary discs that include 
disc winds with observational determinations of the disc lifetime 
and accretion rate in Taurus. 
Using updated estimates for stellar ages in Taurus, 
together with published classifications, we show that the 
evolution of the disc fraction with stellar age 
is similar to that derived for ensembles of stars within 
young clusters. Around 30 percent of stars lose their 
discs within 1~Myr, while the remainder have disc lifetimes 
that are typically in the 1-10~Myr range. We show that the 
latter range of ages is consistent with theoretical 
models for disc evolution, provided that there is a dispersion 
of around 0.5 in the log of the initial disc mass. The 
same range of initial conditions brackets the observed 
variation in the accretion rate of Classical T Tauri stars 
at a given age. We discuss the expected lifetime of discs in 
close binary systems, and show that our models  
predict that the disc lifetime is almost constant for 
separations $d \age 10 \ {\rm au}$. This implies a low 
predicted fraction of binaries that pair a Classical T Tauri 
star with a Weak-lined T Tauri star, and is in better 
agreement with observations of the disc lifetime in binaries 
than disc models that do not include disc mass loss in a wind.
\end{abstract}


\section{Introduction}
Observations of young stars in star forming regions such as 
Taurus Aurigae show that there is a wide dispersion in the 
lifetimes of circumstellar discs, in the range 1-10~Myr
(Strom et al. 1989; Haisch, Lada \& Lada 2001). Although disc 
lifetimes of this order are consistent with the meteoretic evidence 
from our own Solar System (Russell et al. 1996), 
the reason for the large dispersion in the disc lifetime remains unknown. 
The obvious controlling parameter would be the stellar mass, but 
in fact observations of low mass stars in young clusters 
show little correlation between mass and disc 
lifetime (e.g. Rebull et al. 2002).

The simplest disc models also fail to reproduce other properties 
of circumstellar discs. One problem is to match the rapid 
dispersal of the disc. An accretion disc evolves on a single, 
viscous timescale, which in order to reproduce the aforementioned 
disc lifetimes needs to be of the order of $10^6 \ {\rm yr}$. 
Dispersal of the disc on this timescale would lead to the prediction of a 
substantial class of objects with properties intermediate between accreting 
Classical T Tauri stars (CTTS) and discless Weak-lined T Tauri 
stars (WTTS) (Armitage, Clarke \& Tout 1999). In fact, few stars 
are caught in the act of clearing their discs, which constrains the 
disc dispersal time to be short -- of the order of $10^5 \ {\rm yr}$ 
(Simon \& Prato 1995; Wolk \& Walter 1996; Duvert et al. 2000). 

A second problem is that disc evolution is 
largely indifferent to the presence or absence of close binary 
companions. The viscous timescale increases with the characteristic 
radius of the disc, so discs hemmed in 
by close binary companions ought to evolve more rapidly than
discs around relatively isolated stars. Although examples are 
known of binaries that appear to have dissipated their discs at an 
early epoch (Meyer et al. 1997), surveys 
suggest that binary companions as close as 10~au generally have 
little effect on the lifetime or mass accretion rate of the 
disc surrounding the primary (Simon \& Prato 1995; 
Bouvier, Rigaut \& Nadeau 1997; Ghez, White \& Simon 
1997; White \& Ghez 2001). 

Recently, Clarke, Gendrin \& Sotomayor (2001) have shown 
that models which include mass loss from the disc lead to 
a dispersal timescale that is much shorter than the viscous 
evolution timescale, reproducing the two-timescale behavior 
required by the observations. In this paper we 
demonstrate that the same class of models, when combined 
with a modest range of initial disc parameters, are quantitatively 
consistent with the wide dispersion in disc lifetimes. In Section~2 
we set out the disc model and show that when mass
loss is included, the evolution of the accretion rate depends
mainly on the initial disc mass and viscous timescale at large
radius, and only weakly on the radial dependence of disc parameters
such as the surface density and viscosity. In Section~3, we experiment
with the range of initial conditions that are required to reproduce
the observed age distributions of CTTS and WTTS in Taurus-Aurigae,
and show that this range also reproduces the observed scatter in
accretion rates in CTTS inferred by Gullbring et. al. (1998).  
In Section~4, we apply the model to binary star systems, and show
that due to the wind  mass loss from large radii in the disc, the
disc lifetime may {\it increase} with decreasing disc binary separation
over the binary separation range $10-100$~au. Section~5 summarizes
our conclusions.

\section{Disc model}

\subsection{Evolution equation}

The evolution of the surface density, $\Sigma (r,t)$, of a geometrically 
thin accretion disc is described by (Lynden-Bell \& Pringle 1974; Pringle 1981),
\begin{equation}
 { {\partial \Sigma} \over {\partial t} } = 
 { 3 \over r } { \partial \over {\partial r} } 
 \left[ r^{1/2} { \partial \over {\partial r} } 
 \left( \nu \Sigma r^{1/2} \right) \right] + 
 \dot{\Sigma}_{\rm wind} \left( t \right), 
\label{eq_evolve}
\end{equation}
where the term $\dot{\Sigma}_{\rm wind}$ represents 
the rate of mass loss per unit surface area of the disc in the 
equatorial plane. We assume that any mass lost leaves the disc 
with Keplerian specific angular momentum, and thus does not 
drive radial inflow or outflow in the remaining disc gas. For 
the kinematic viscosity $\nu$ we adopt a simple power-law 
in radius, 
\begin{equation} 
 \nu = \nu_0 r^{\beta},
\label{eq_nu}
\end{equation}
with $\nu_0$ and $\beta$ constants. 
Within the context of $\alpha$ disc models (Shakura \& Sunyaev 1973), 
the viscosity in a protoplanetary disc is determined by the intrinsic 
efficiency of angular momentum transport (i.e. the value of $\alpha$); 
by the relative contribution of viscous heating and stellar irradiation;  
and by the opacity which controls the vertical disc structure. A 
passive irradiated disc, in which both the central temperature and 
the surface temperature decline with radius as $T_c \propto r^{-1/2}$, 
has a viscosity (assuming a constant $\alpha$) which scales as 
$\nu \propto r$. More elaborate calculations of protoplanetary 
disc structure -- though still within the $\alpha$ disc formalism -- 
likewise yield $\beta \sim 1$ (e.g. Bell et al. 1997; 
D'Alessio et al. 1998). Only limited observational constraints are 
available at the (small) radii of interest, though $\beta = 1$ 
is consistent with VLA 
observations of TW~Hydrae by Wilner et al. (2000). 

In a steady-state, the surface density profile corresponding 
to a given mass accretion rate $\dot{M}$ is,
\begin{equation} 
 \nu \Sigma = { \dot{M} \over {3 \pi} } 
 \left( 1 - \sqrt{ {r_{\rm in} \over r } } \right),
\label{eq_steady}
\end{equation} 
where we have assumed standard zero-torque boundary conditions 
at $r_{\rm in}$, the inner edge of the disc.

Equation (\ref{eq_evolve}) is a diffusion equation for the 
surface density. Provided that the surface density evolution 
is controlled by angular momentum transport rather than 
mass loss, the characteristic timescale for evolution is 
the viscous time,
\begin{equation} 
 t_\nu = { r^2 \over \nu }.
\label{eq_timescale}
\end{equation}   
Since the viscous timescale increases with radius, the 
evolution of a disc described by equation (\ref{eq_evolve}) 
speeds up at late times if the outer edge of the disc 
is truncated by mass loss.

Standard numerical methods (e.g. Pringle, Verbunt \& Wade 1986) 
are employed to solve equation (\ref{eq_evolve}) on a non-uniform 
radial mesh. We use zero-torque ($\Sigma(r_{\rm in}) = 0$) boundary 
conditions at the inner edge of the disc. The spatial domain is 
large enough that the outer boundary condition has no 
influence upon the results. Formally, we set the radial 
velocity $v_r = 0$  at this boundary. 

\subsection{Disc mass loss}

Observations provide direct evidence for the existence of two 
types of disc outflow, jets and photoevaporative flows. Of these, 
photoevaporative flows, which lead to mass loss from the outer 
regions of the disc, are the most obviously promising for producing 
two-timescale disc evolution and rapid disc dispersal (Clarke, 
Gendrin \& Sotomayor 2001; Matsuyama, Johnstone \& Hartmann 2002).  
We consider mass loss via photoevaporation exclusively in this paper, 
while noting that almost identical behaviour is also possible in models 
of magnetically driven outflows in which mass loss extends beyond the 
inner few au of the disc (e.g. Ostriker 1997).

Photoevaporative flows are thermal winds that occur when the 
disc surface is heated by exposure to ultraviolet radiation 
(Hollenbach et al. 1994; Johnstone, Hollenbach \& Bally 1998; 
Hollenbach, Yorke \& Johnstone 2000; Richling \& Yorke 2000).
The heated gas can escape at radii $r > r_g$, where the critical 
radius $r_g$ is given (approximately) by the condition that the sound 
speed $c_s$ of the hot gas exceeds the local escape velocity,
\begin{equation} 
 r_g \sim { {GM} \over c_s^2 }.
\label{eq_escaperadius}
\end{equation} 
For parameters appropriate to T Tauri stars, $r_g \sim 5-10 \ {\rm au}$. 
The flux of photoevaporating radiation can be external, as in  
Orion where UV radiation from massive stars is responsible 
(Churchwell et al. 1997; Johnstone, Hollenbach \& Bally 1998; 
Scally \& Clarke 2001; Henney et al. 2002). Alternatively, a 
lower level of mass loss can result from the UV and X-ray flux 
of the low mass star itself (Shu, Johnstone \& Hollenbach 1993).

In Taurus, there are no massive stars which could generate a strong 
external flux of photoevaporating radiation. We therefore adopt a 
simple mass loss prescription appropriate for the case where the 
mass loss is driven by local irradiation (Hollenbach et al. 1994). 
The mass loss per unit area of the disc, 
$\dot{\Sigma}_{\rm wind}$, is a function of radius but not of 
time\footnote{Implicitly, this assumes that the photoevaporating 
flux is not tied to the accretion rate. Models by 
Matsuyama, Johnstone \& Hartmann (2002) show that photoevaporation
driven solely by the star-disc accretion shock fails to disperse 
the disc quickly enough.}, and is given by,
\begin{eqnarray}
 \dot{\Sigma}_{\rm wind} & = & 0, \,\,\,\,\,\,\,\,\,\,\, r < r_g \nonumber \\
 \dot{\Sigma}_{\rm wind} & \propto & r^{-5/2}, \,\,\,\,\, r > r_g.
\label{eq_massloss}
\end{eqnarray}  
We express the normalization of the mass loss via the parameter $\dot{M}_{\rm wind}$, 
which is defined as the total mass loss rate if the disc extends to 25~au. The 
instantaneous mass loss rate will therefore differ from this depending upon 
the outer disc radius. We take $r_g = 5 \ {\rm au}$.

\subsection{Model parameters}

\begin{figure}
\psfig{figure=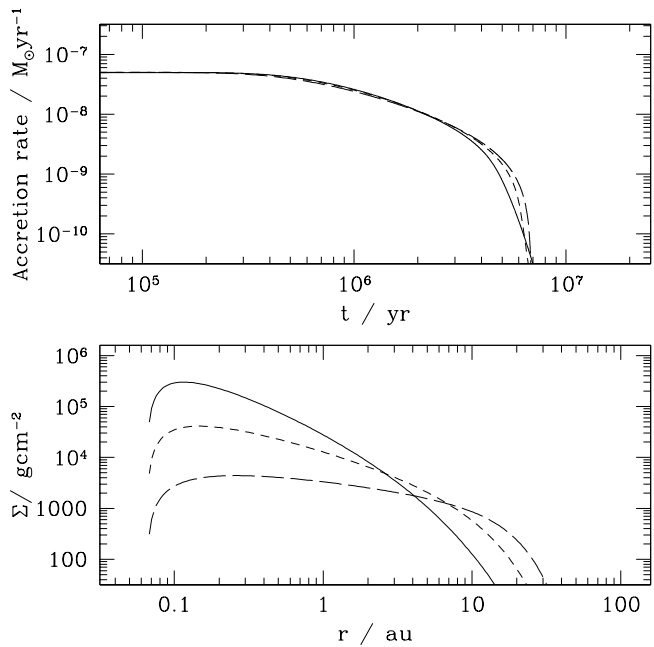,width=3.5truein,height=3.5truein}
\caption{The accretion rate (upper panel) and the surface density profile 
	(lower panel, plotted at $t = 0.5 \ {\rm Myr}$) for disc models 
	with $\beta = 3/2$ (solid line), $\beta = 1$ (short dashes), and 
	$\beta = 1/2$ (long dashes). The other parameters of the models 
	are as given in Table~1. Models with very different viscosity 
	laws and surface density profiles can yield almost identical 
	accretion rate histories.}
\label{fig_degeneracy}
\end{figure}

\begin{table*}
\begin{tabular}{cccc}
	 & $\beta = 1/2$ & $\beta = 1$ & $\beta = 3/2$ \\ \hline
$\nu (r=1 \ {\rm au})$ & $6.5 \times 10^{13} \ {\rm cm}^2 {\rm s}^{-1} $ & 
	$7.5 \times 10^{13} \ {\rm cm}^2 {\rm s}^{-1} $ & 
	$1.3 \times 10^{14} \ {\rm cm}^2 {\rm s}^{-1} $ \\
Initial disc mass & $0.15 \ M_\odot$ & 	$0.125 \ M_\odot$ & $0.10 \ M_\odot$ \\ 
Disc mass at 2~Myr & $0.07 \ M_\odot$ & $0.045 \ M_\odot$ & $0.027 \ M_\odot$ \\
Viscous time at 10~au & 3.9~Myr & 4.8~Myr & 3.9~Myr \\
Inner radius for disc mass loss & 5~au & 5~au & 5~au \\	
$\dot{M}_{\rm init}$ & $5 \times 10^{-8} \ M_\odot {\rm yr}^{-1}$ & 
	$5 \times 10^{-8} \ M_\odot {\rm yr}^{-1}$ & $5 \times 10^{-8} \ M_\odot {\rm yr}^{-1}$ \\
$\dot{M}_{\rm wind}$ & $5 \times 10^{-9} \ M_\odot {\rm yr}^{-1}$ & 
	$5 \times 10^{-9} \ M_\odot {\rm yr}^{-1}$ & $5 \times 10^{-9} \ M_\odot {\rm yr}^{-1}$ \\ \hline
\label{models_table}
\end{tabular}      
\caption{Summary of parameters for the three models shown in Fig.~1. All of these 
         models are broadly consistent with the observed evolution of the 
	 accretion rate and disc mass in young clusters such as Taurus.}
\end{table*}

The model described above is defined by two parameters describing the 
disc physics ($\nu_0$ and $\beta$), two describing the wind
($\dot{M}_{\rm wind}$ and $r_g$), and two for the initial 
conditions ($\dot{M}_{\rm init}$ and the initial disc mass)\footnote{Equivalently, 
we could parameterize the initial conditions in terms of the initial mass 
and angular momentum content of the disc.}.
Sensible values for most of these parameters can be estimated 
from existing observations. We adopt $\dot{M}_{\rm init} = 5 \times 10^{-8} \ 
M_\odot {\rm yr}^{-1}$, since most T Tauri stars have accretion rates 
of this order or lower (Gullbring et al. 1998). With this initial 
accretion rate, a disc lifetime of 1-10~Myr mandates an 
initial disc mass of $\approx 0.1 M_\odot$. After 2~Myr of 
evolution, this initial disc mass has declined (see Table~1) to 
a few hundredths of a Solar mass, a value which is broadly consistent 
with estimates of T Tauri disc masses (Beckwith \& 
Sargent 1991). The assumption that mass loss is driven by 
photoevaporation fixes $r_g$ to be 5-10~au, and the requirement 
that this mass loss drives rapid dispersal of the disc suggests 
that $\dot{M}_{\rm wind}$ be no more than a couple of orders 
of magnitude smaller than $\dot{M}_{\rm init}$ (Clarke, Gendrin \& 
Sotomayor 2001).

Determining the parameters for the disc viscosity is more difficult. 
We have already noted that theoretical models for protoplanetary 
discs favour values for $\beta$ around unity, but no-one would 
claim that this parameter is as yet accurately determined. 
We have therefore chosen for our fiducial models {\em combinations} of $\nu_0$ and 
$\beta$, shown in Table~1 along with the other parameters of the disc model, 
that yield a disc lifetime of approximately 5~Myr. This choice of  
disc lifetime fixes the viscous time at a radius of around 10~au, but 
does not constrain either $\nu_0$ or $\beta$ individually. Indeed, 
as shown in Figure~1, disc models that include a disc wind have only a limited 
period during which the accretion rate drops as a power-law defined by 
the value of $\beta$ in the viscosity law. As a consequence of this, 
models with different choices of $\beta$ can be constructed that 
yield almost identical evolution of the disc accretion rate, even 
though they have entirely different surface density profiles. This 
degeneracy implies that it is not possible to discriminate observationally -- even 
if we had perfect data -- between models with different $\beta$, using 
only measurements of the evolution of the accretion rate with time. 
We cannot, therefore, use the accretion rate evolution measured by 
Gullbring et al. (1998) to constrain our models in the same way as 
was done for disc models without mass loss by  Hartmann et al. (1998) 
and Stepinski (1998). In the 
following discussion, we will leave $\beta$ as an unconstrained parameter, 
and consider the three models outlined in Table~1 as equally valid 
possible descriptions of the disc.

\begin{figure*}
\psfig{figure=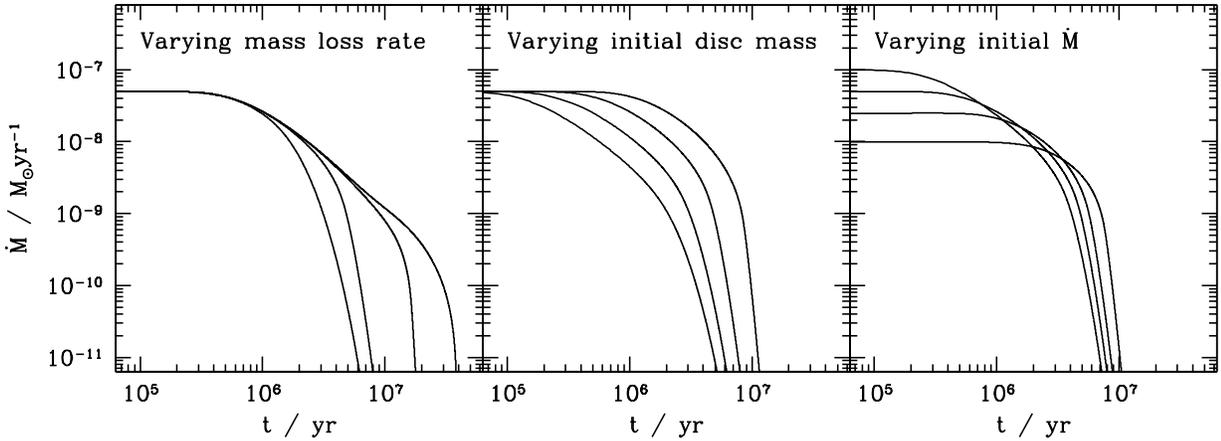,width=7.0truein,height=7.0truein}
\vspace{-3.7truein}
\caption{Illustration of how the accretion rate evolution depends upon the 
	model parameters. The fiducial model has $\nu \propto r^{3/2}$, an 
	initial disc mass $M_{\rm disc} = 0.1 \ M_\odot$, $\dot{M}_{\rm init} = 
	5 \times 10^{-8} \ M_\odot {\rm yr}^{-1}$ and $\dot{M}_{\rm wind} = 
	5 \times 10^{-9} \ M_\odot {\rm yr}^{-1}$. We adopt $r_{\rm in} = 0.066 \ 
	{\rm au}$, and $r_{\rm out} = 100 \ {\rm au}$. The leftmost panel shows models 
	with mass loss rates $\dot{M}_{\rm wind} = 5 \times 10^{-11}, 5 \times 10^{-10}, 
	5 \times 10^{-9}, 5 \times 10^{-8} \ M_\odot {\rm yr}^{-1}$. The centre 
	panel shows the effect of changing the initial disc mass to $0.025, 0.05, 
	0.1, 0.2 \ M_\odot$. The rightmost panel shows models with $\dot{M}_{\rm init} = 
	10^{-8}, 2.5 \times 10^{-8}, 5 \times 10^{-8}, 10^{-7}  \ M_\odot {\rm yr}^{-1}$.
	Rapid dispersal of the gas is a generic feature of all of these models.}
\label{fig_variants}
\end{figure*}

To illustrate how the results depend upon some of the other model 
parameters, we show in Fig.~\ref{fig_variants} how the accretion 
rate evolves with time for different choices of the initial disc 
mass, initial accretion rate, and disc mass loss rate. 
We start with a disc of mass $M_{\rm disc}$, and accretion rate $\dot{M}_{\rm init}$. 
The disc has an initial surface density profile defined by equation (\ref{eq_steady}) out to 
the radius where the enclosed mass is $M_{\rm disc}$, and $\Sigma = 0$ at greater 
radii. As is obvious from Fig.~\ref{fig_variants}, the two-timescale behaviour 
discussed by Clarke, Gendrin \& Sotomayor (2001) occurs in a 
range of disc models which incorporate mass loss.
The decline of the accretion rate steepens dramatically in all 
models at the epoch when mass loss in the disc wind becomes important. 

\section{The disc fraction in Taurus}

In a recent study, Haisch, Lada \& Lada (2001) determined (using 
{\it JHKL} band photometric surveys)the fraction
of stars possessing discs as a function of the previously published 
mean cluster age. They found an evolution in the disc fraction shown 
in the lower panel of Figure~4. However, since the 
spread in stellar ages can be comparable to the mean age for 
clusters that are only $\sim 10^6$~yr old, it is not immediately 
obvious whether the smooth decline in disc fraction with `age' 
seen in the Haisch, Lada \& Lada (2001) sample reflects primarily,
\begin{itemize}
\item
A true dispersion in the lifetime of circumstellar discs.
\item
A convolution of a constant disc lifetime with an 
extended epoch of star formation in each cluster.
\end{itemize}
To distinguish between these possibilities, it is 
necessary to date the stars 
within young clusters on a star by star basis. We 
present such an analysis below, and show that in 
Taurus the evidence suggests that for low mass stars 
there {\em is} a significant 
intrinsic dispersion in disc lifetime. 

\subsection{Data}

To study the evolution of the disc fraction with stellar age, 
we make use of data compiled for a study of star formation 
in Taurus by Palla \& Stahler (2002). 
The sample of 151 stars includes 
approximately 10 stars from Wichmann et al. (2000) whose 
pre-main-sequence status is confirmed by lithium observations, 
and which fall within the boundaries of the $^{12}$CO map that 
contains the bulk of the Taurus-Auriga population. ROSAT All Sky 
Survey sources that fall at larger distances have not been 
included. Of this sample, 67 stars are classified as CTTS, 
and 57 are WTTS. For 27 stars discovered in more recent surveys 
we were unable to find published classifications, and these 
are excluded from further analysis.

For each of the stars in the CTTS and WTTS samples, we have 
estimated ages using the evolutionary tracks of Palla \& Stahler (1999). 
In general, age estimates for pre-main-sequence stars are 
subject to considerable uncertainty, due to different assumptions 
as to the initial conditions and accretion history (Tout, Livio \& 
Bonnell 1999; Hartmann 2001; Baraffe et al. 2002). In the 
specific case of Taurus, Hartmann (2003) has argued that the 
ages of the higher mass stars (with masses $\age 1 \ M_\odot$) 
are especially uncertain, due to 
different treatments of the stellar birthline (e.g. Hartmann, 
Cassen \& Kenyon 1997; Palla \& Stahler 1999) and possible 
foreground contamination. We therefore further restrict 
the sample under consideration to those stars with 
effective temperatures $T_e \leq 4350 \ {\rm K}$, leaving 
a final sample of 57 CTTS and 41 WTTS. With the effective 
temperature cut, these are all low mass stars.

\begin{figure}
\psfig{figure=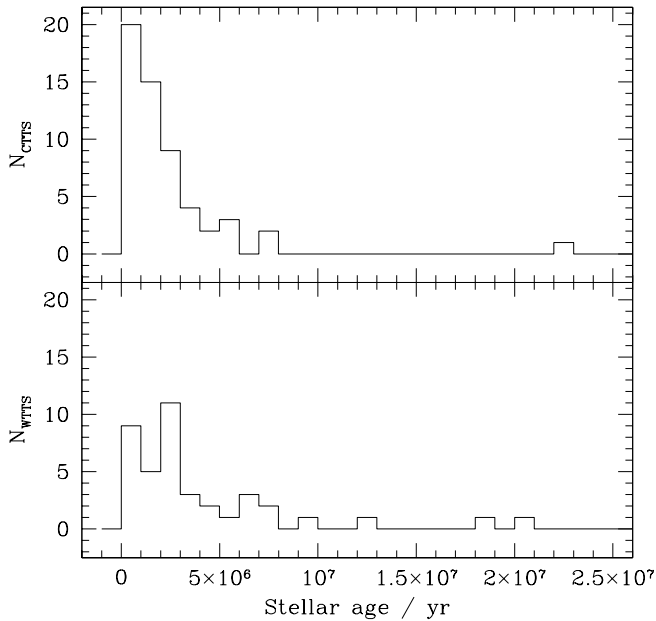,width=3.5truein,height=3.5truein}
\caption{Distribution of the ages of stars classified as 
	Classical T Tauri stars (upper panel) and Weak-Lined 
	T Tauri stars (lower panel) in Taurus. Only stars 
	with effective temperature $T_e \leq 4350 \ {\rm K}$ 
	are included in the histograms.}
\label{fig_ages}
\end{figure}

Figure~\ref{fig_ages} shows histograms of the age distribution 
of the low mass CTTS and WTTS in Taurus. The most immediately 
obvious conclusion to be drawn from the Figure is that there 
is substantial overlap in the derived ages of the CTTS and
the WTTS. However, there is also evidence that  
the WTTS are significantly older as a class. A
KS test reveals that the age distributions of the WTTS and CTTS samples
differ at more than the 2$\sigma$ level (the probability that the distributions 
are the same is approximately 0.03). 

Although the empirical conclusion that the WTTS are significantly 
older than the CTTS falls short of statistical proof, the 
evidence is consistent with our expectations if CTTS turn into WTTS. 
Indeed, if no such difference
were found, one would either have to posit no evolutionary link
between the two (thus begging the question of what CTTS {\it should}
turn into) or else appeal to contrived models in which the lifetime
of discs were shorter in the case of stars born more recently. We shall
show below that the Taurus data is instead compatible with the simple
hypothesis that CTTS turn into WTTS, but that the clock for this
process varies from star to star. In Chamaeleon I, by contrast,
it has been claimed that the age distribution of CTTS and WTTS is
indistinguishable (Lawson, Feigelson \& Huenemorder 1996). It is
now known, however, that this dataset was missing a number of young 
CTTS (Persi et al 2000), and a more complete census of this region
may not ultimately yield an answer different from the Taurus result
reported here.

\begin{figure}
\psfig{figure=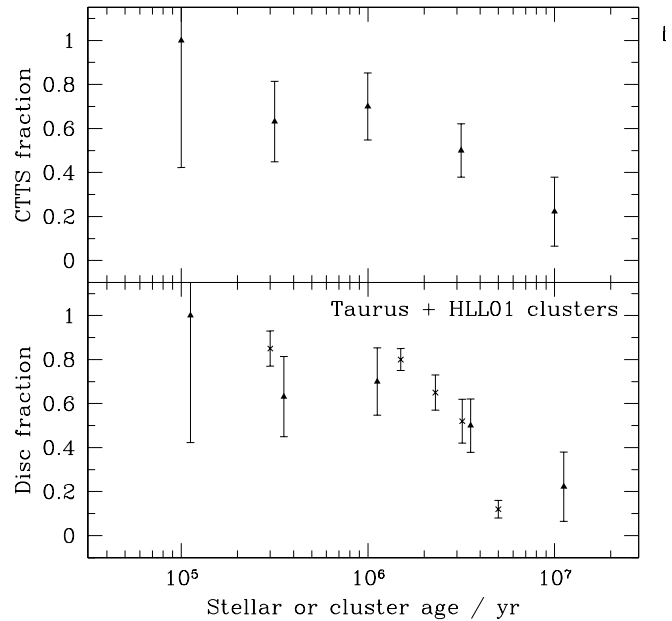,width=3.5truein,height=3.5truein}
\caption{The disc fraction as a function of age in Taurus and 
	other young clusters. The upper panel shows the Classical 
	T Tauri fraction $f_{\rm CTTS} = N_{\rm CTTS} / (N_{\rm CTTS} + N_{\rm WTTS})$ 
	as a function of stellar age, using published classifications 
	and our updated estimates for the ages. The lower panel 
	compares the Taurus data (triangles, shifted slightly to the 
	right for clarity) to the evolution of the disc 
	fraction measured by Haisch, Lada \& Lada (2001) for young 
	clusters (points).}
\label{fig_data}
\end{figure}

Figure~\ref{fig_data} shows the CTTS fraction in Taurus as a function 
of stellar age. We have binned the data into 5 bins, evenly 
spaced in the $\log$ of the estimated stellar age. Most of the 
transition between CTTS and WTTS occurs for ages between 1~Myr and 
10~Myr. At the youngest inferred ages, CTTS are dominant, but 
there is still a significant admixture of WTTS. We 
note, however, that for such young ages -- less than a Myr -- 
theoretical uncertainties in the age are probably comparable 
to the age itself (Baraffe et al. 2002). Any conclusions we 
might draw as to the evolution of the CTTS fraction prior 
to 1~Myr should therefore be treated with caution.

Figure~\ref{fig_data} also shows the evolution of the disc 
fraction as measured by Haisch, Lada \& Lada (2001) in a 
sample of 6 young clusters with estimated ages ranging 
between 0.3~Myr and 30~Myr. Within the errors, which are 
substantial (arising both from the limited number of 
stars, and from the uncertainty in cluster age which 
is not shown in the Figure), the same trend with age 
is seen as for the Taurus data.

\subsection{A model for the dispersion in disc lifetime}

\begin{figure}
\psfig{figure=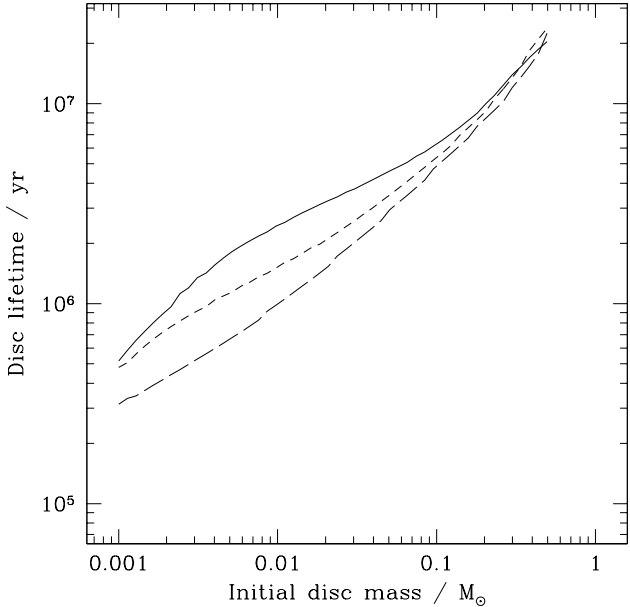,width=3.5truein,height=3.5truein}
\caption{The disc lifetime as a function of the initial disc mass, for 
	models with $\beta = 3/2$ (solid line), $\beta = 1$ (short dashes), 
	and $\beta = 1/2$ (long dashes). The 
	initial accretion rate is assumed constant for all models.}
\label{fig_lifetime}
\end{figure}

\begin{figure}
\psfig{figure=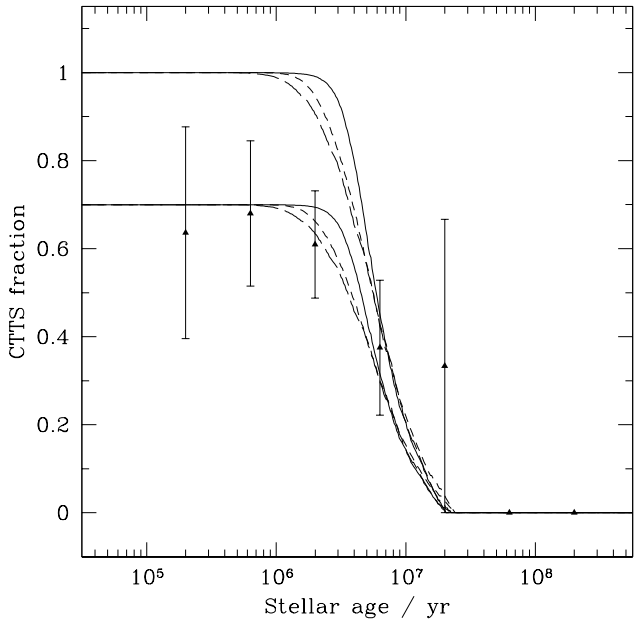,width=3.5truein,height=3.5truein}
\caption{Evolution of the disc fraction predicted by models with $\beta = 3/2$ 
	(upper solid line), $\beta = 1$ (short dashes), and $\beta = 1/2$ (long 
	dashes), assuming in each case that there is a dispersion of 0.5 in the 
	log of the initial disc mass. Better, but {\em ad hoc} fits to the 
	data (lower curves), are obtained 
	by assuming that $\approx 30 \%$ of stars lose their discs via other 
	processes within the first Myr.}
\label{fig_cluster}
\end{figure}

A dispersion in the disc lifetime could be caused by several factors, 
including different disc initial conditions or varying rates of mass 
loss. For photoevaporation, the mass loss rate is predicted to vary 
only weakly (as the square root) with the ionizing flux. We have 
therefore explored whether the data shown in Fig.~\ref{fig_data} 
can instead be explained as arising from different initial conditions 
for the discs. We focus on the effect of different initial disc 
masses, as Fig.~\ref{fig_variants} indicates that the disc lifetime 
changes markedly as we vary this parameter.

Fig.~\ref{fig_lifetime} shows how the disc lifetime varies with the 
initial disc mass. We define the disc lifetime as the time at which 
the accretion rate first drops below $10^{-10} \ M_\odot {\rm yr}^{-1}$ 
(once this epoch is reached, as shown in Figure~1, mass loss in the 
wind dominates the evolution, and the accretion rate is dropping 
precipitously). Apart from changing $M_{\rm disc}$, the other parameters 
are kept fixed at the values shown in Table~1 (i.e. we change the initial 
conditions for the disc, but keep the parameters describing the disc 
viscosity and disc wind constant). For all three disc models, the disc 
lifetime has an approximately power-law dependence upon the initial 
disc mass, with an index of $\approx 0.5$. 

Figure~\ref{fig_cluster} shows the predicted disc fraction as 
a function of age, under the assumption that $\log (M_{\rm disc})$ is 
distributed as a Gaussian around the central values given in 
Table~1. We find that the general trend in the plot 
of disc fraction with age can be reproduced provided that there is 
a 1 sigma dispersion of approximately 0.5 in the $\log$ of the initial 
disc mass. A dispersion in initial disc mass of this magnitude 
leads to almost all stars losing their discs between 1~Myr 
and 10~Myr, as required to fit the evolution of the CTTS fraction 
in Taurus, and the evolution of the disc fraction in the 
clusters studied by Haisch, Lada \& Lada (2001).

Although the models do a reasonable job in reproducing the width of the 
transition seen in Fig.~\ref{fig_cluster}, they fail in that they 
predict a 100\% disc fraction for the youngest stars, those with 
ages less than a Myr. Indeed, from Fig.~\ref{fig_lifetime} we find 
that a disc lifetime as short as (say) 0.5~Myr requires an initial 
disc mass that is very small -- of the order of 
few~$\times 10^{-3} \ M_\odot$. As noted previously, one possibility 
is simply that the ages of the youngest WTTS have 
been underestimated. We find, however, that the spatial distribution 
of these young WTTS, relative to the cluster gas, is more consistent with 
the distribution of the young CTTS (Palla \& Stahler 2002) than of the 
older stars.  Another possibility is that these objects, classified as WTTS
on the grounds of their equivalent width in $H_\alpha$, have  highly
variable line emission, and have thus been classed as WTTS even though
they are not {\em bona fide} discless objects. However, we find that
these systems are also lacking significant excesses both in the near
and far infrared, thus confirming their discless status. These
considerations support the contention that the apparently young,
discless stars really are young and discless, and constitute a separate 
population which have lost their discs at a earlier epoch due to 
processes that do not affect the majority of stars. An {\em ad hoc} model 
in which 30\% of stars lose their discs within a Myr, while the rest 
evolve in the way which we have modelled, is shown in Fig.~\ref{fig_cluster}, 
and is consistent with the available observations. 

If stars form in isolation, then it is hard to understand why a 
significant fraction of them should lose their discs very early. 
Rapid early accretion onto the star is expected to cease when the 
disc first becomes gravitationally stable, at a mass of the 
order of $0.1 \ M_*$, after which time slower `viscous' evolution 
continues. Within this framework, initial disc masses that are 
smaller than average by one to two orders of magnitude are 
puzzling. In an environment where stars
are strongly sub-clustered at birth (Scally \& Clarke 2002), however, 
close encounters between stars can efficiently reduce the disc mass, both 
directly by unbinding disc material (Clarke \& Pringle 1993), 
and indirectly by reducing the outer disc radius and hastening 
viscous evolution (Armitage \& Clarke 1997). Recent simulations 
of clustered star formation (Bate, Bonnell \& Bromm 2003) show 
that the required interactions may well be extremely common. In the 
Bate et al. (2003) calculation, {\em most stars} have encounters 
with another star with a minimum separation distance of a few 
au or smaller. The young WTTS seen in Taurus may simply be 
the fraction of those stars which fail to accrete a significant 
new disc following their last close flyby.

\subsection{Evolution of the accretion rate of CTTS}

\begin{figure}
\psfig{figure=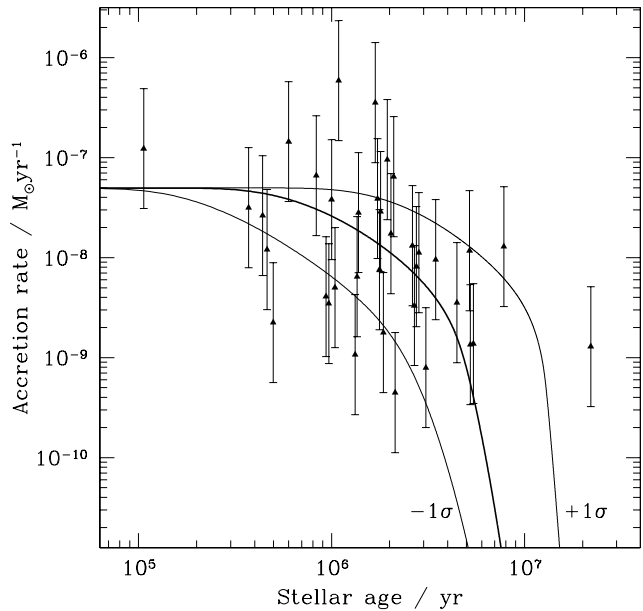,width=3.5truein,height=3.5truein}
\caption{The accretion rate of Classical T Tauri stars in Taurus, as 
	estimated by Hartmann et al. (1998), plotted against the 
	stellar age. The heavy central curve shows the predicted 
	evolution of the accretion rate for the $\beta = 3/2$ disc 
	model outlined in Table~1. The upper and lower curves 
	show the accretion rate for models in which the initial 
	disc mass has been increased or decreased by a factor 
	of three.}
\label{fig_accrate}
\end{figure}

With the caveats noted above, the evolution of the disc fraction 
in Taurus is consistent with a theoretical model in which 
differences in disc lifetime are ascribed to differences in 
the initial disc mass. A $1 \sigma$ dispersion of around 0.5 
in the $\log$ of the initial disc mass is required to reproduce 
typical disc lifetimes of 1-10~Myr. Figure~\ref{fig_accrate} 
shows how this model compares with observations of the accretion 
rate of CTTS in Taurus. We plot estimates of $\dot{M}$ by 
Gullbring et al. (1998) (as tabulated in Hartmann et al. 1998) 
against our own revised determinations of the stellar ages. The 
vertical error bars show Hartmann et al.'s (1998) best estimate 
of the error in determining the accretion rate, which is 0.6 in 
$\log(\dot{M})$. Against this data, we plot theoretical 
curves for the $\beta = 3/2$ disc model. We show the predicted 
evolution of $\dot{M}$ for the central estimate of the initial 
disc mass of $0.1 \ M_\odot$, and for the models with initial 
disc masses differing by $\pm 1 \sigma$.

From the Figure, it is clear that the $\pm 1 \sigma$ curves 
bracket the observed range of CTTS accretion rates at a 
given age. The dispersion in initial disc mass required 
to reproduce the observed range in disc lifetimes also 
produces roughly the correct spread in accretion rates. 
Since we can reproduce the same result with other 
choices of $\beta$, our interpretation is that the Hartmann et al. (1998) 
plot of $\dot{M}$ against $t$, shown in a modified form as 
Figure~\ref{fig_accrate}, tells us more about the variation 
in the initial conditions for discs than it does about the 
internal physics governing their evolution. 

\section{The influence of binary companions}

\begin{figure}
\psfig{figure=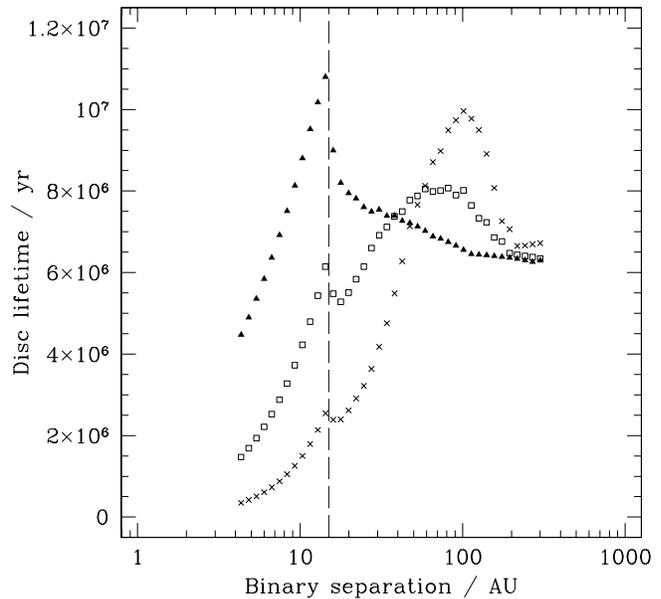,width=3.5truein,height=3.5truein}
\caption{The predicted disc lifetime in binary systems, for models 
	with $\beta = 3/2$ (filled triangles), $\beta = 1$ (open squares), 
	and $\beta = 1/2$ (crosses). We assume that the disc is truncated 
	at $1/3$ of the binary separation, and that there is no replenishment 
	of the circumstellar discs from circumbinary material. The vertical 
	dashed line shows the separation below which the discs suffer no 
	mass loss.}
\label{fig_binary}
\end{figure}

Most T Tauri stars are found in binary systems (Leinert et al. 1993; Ghez, 
Neugebauer \& Matthews 1993; Simon et al. 1995), which can affect the 
evolution of circumstellar discs in several ways. First, gravitational 
torques from the companion act to truncate the disc (Paczynski 1977), 
and prevent it from expanding freely. Second, if the binary is sufficiently 
close, it will limit the reservoir of gas available to be accreted. Both 
of these effects would be expected to reduce the lifetime of discs in 
binary systems as compared to single stars. A substantially reduced disc 
lifetime in binaries is not, however, observed (Simon \& Prato 1995; 
Bouvier, Rigaut \& Nadeau 1997; Ghez, White \& Simon 1997; White \& 
Ghez 2001).

Fig.~\ref{fig_binary} shows how binaries alter the predicted disc 
lifetime in our models. We have assumed that the only effect of the 
binary on the initial surface density distribution is to truncate 
the disc at a radius $r_t = d/3$, where $d$ is the binary separation. 
The initial accretion rate in the binary models is therefore the 
same as for the isolated disc models. At $r_t$, we model the 
torques from the companion using a $v_r = 0$ boundary condition.
Parameters for the disc wind and the disc viscosity remain 
unchanged.

By construction, the disc lifetime for sufficiently wide binaries 
tends to the observed value of $\approx 6 \ {\rm Myr}$. At smaller 
separations, all three disc models display similar behaviour. The 
disc lifetime initially {\em increases}, reaching a peak at a 
separation of 15~au to 100~au, depending upon the model. This 
increase occurs because the disc, which is prevented from expanding 
freely, has less area from which mass can be lost in a wind. This 
compensates for, and initially overwhelms, the tendency for smaller 
disc to have reduced viscous timescales. As a result, only very 
close binaries ($d$ less than 10~au to 40~au, depending upon the 
model) are predicted to lead to a reduction in the 
disc lifetime as compared to the lifetime of an isolated disc.

In addition to asking whether the disc lifetime around the primary 
is a function of binary separation, we can also ask whether 
differential evolution of the primary and secondary disc is 
expected to yield `mixed' binary systems that pair CTTS and 
WTTS. The naive expectation is that the lower mass component 
of the binary, whose Roche lobe encloses a smaller disc, should 
lose its disc and become a WTTS first. Observations, however, show that 
mixed pairs are rather rare. Using a sample of close binaries (most 
of which had projected separations between 15~au and 100~au) Hartigan 
\& Kenyon (2002) found that only 25~percent (4 out of 16 systems) 
of binaries containing at least one CTTS were mixed systems. Combining 
these results with earlier studies (Brandner \& Zinnecker 1997; 
Prato \& Simon 1997; Duchene et al. 1999), which obtained similarly 
small mixed fractions, there is no evidence for a trend in the 
fraction of mixed systems with separation. Comparing the observations to 
disc models that do not include winds, this is only consistent with 
theoretical expectations if the initial conditions yield more massive 
discs around the lower mass star (Armitage, Clarke \& Tout 1999).

If we assume that the initial disc accretion rate, and the mass loss rate, 
are the same for both stars in a binary, then the disc models 
presented in this paper predict a low level of mixed WTTS / CTTS 
binaries. For a mass ratio $q=0.5$, the maximum radii of discs 
surrounding the primary and secondary differ by around 50 percent 
(Papaloizou \& Pringle 1977). From Figure~\ref{fig_binary}, we 
then estimate that an initially doubly strong (CTTS + CTTS) 
binary with separation $d \age 10$~au will appear as a mixed 
binary for at most 10\% (if $\beta=3/2$) to 20\% (if $\beta=1$) 
of the time. A larger fraction (up to 50\%) of mixed binaries 
is predicted for the $\beta=1/2$ disc model. 
As was the case with the disc lifetime, therefore, only very 
close binaries or binaries with extreme mass ratios are expected to show 
significant differential disc evolution. 

\section{Summary}
We have shown that protoplanetary disc models which 
include mass loss from the disc at large radius are 
generally consistent with observations of accretion 
in T Tauri stars. In particular, we 
can simultaneously reproduce both the rapid disc dispersal 
required by observations (Simon \& Prato 1995; 
Wolk \& Walter 1996; Duvert et al. 2000), and the 
near-independence of the disc lifetime on the 
binary environment (White \& Ghez 2001). The 
evolution of the disc fraction in young clusters, 
which we have studied in detail for Taurus, can also 
be modelled within this framework, provided that 
we allow for a spread in initial disc masses 
around a central value of $\approx 0.1 \ M_\odot$.
We believe that the required level of variation in the 
initial conditions of discs is a plausible outcome 
of the turbulent conditions that give rise to star 
formation within clusters. Simulations of clustered 
star formation (e.g. Bate, Bonnell \& Bromm 2003) 
show that interactions between small groups of forming 
stars and their discs are almost inevitable. In 
such a dynamic environment, we expect large star to 
star variations in the initial mass and angular 
momentum content of circumstellar material. In fact, a 
significant fraction of the stars formed in the 
Bate et al. (2003) simulation suffer encounters 
with other stars that are close enough to truncate 
circumstellar discs to sizes of au or smaller. Such 
interactions provide a possible origin for the 
otherwise puzzling observation of a population of 
apparently very young Weak lined T Tauri stars seen in the 
Taurus region. 

\section*{Acknowledgements}

This work has been supported by the European Community's Research Training
Network under contract HPRN-CT-2000-0155, ``Young Stellar Clusters''. We 
thank the referee, Lee Hartmann, for a prompt and valuable review. 
CJC gratefully acknowledges support from the Leverhulme Trust in the
form of a Philip Leverhulme Prize, and from JILA via their 
visitors' programme. PJA thanks the Institute of Astronomy for 
support from their visitors' programme.

\end{document}